\documentclass[]{emulateapj}

\def\kms{\,{\rm {km\, s^{-1}}}}

\shorttitle{The First Hypervelocity Star from the LAMOST Survey}
\shortauthors{Zheng et al.}

\begin{document}

\title{The First Hypervelocity Star from the LAMOST Survey}

\author{Zheng Zheng$^1$, Jeffrey L. Carlin$^2$, Timothy C. Beers$^3$, 
        Licai Deng$^4$,  Carl J. Grillmair$^5$, Puragra Guhathakurta$^6$, 
        S\'ebastien L\'epine$^7$, Heidi Jo Newberg$^2$, Brian Yanny$^8$, 
        Haotong Zhang$^4$, Chao Liu$^4$, Ge Jin$^9$, and Yong Zhang$^{10}$}
\affil{
$^1$Department of Physics and Astronomy, University of Utah, UT 84112, USA; zhengzheng@astro.utah.edu
}
\affil{
$^2$Department of Physics, Applied Physics, and Astronomy, Rensselaer Polytechnic Institute, 110 8th Street, Troy, NY 12180, USA
}
\affil{
$^3$National Optical Astronomy Observatory, Tucson, AZ 85719, USA, and JINA: Joint Institute for Nuclear Astrophysics
}
\affil{
$^4$National Astronomical Observatories, Chinese Academy of Sciences, Beijing 100012, China
}
\affil{
$^5$Spitzer Science Center, 1200 E. California Blvd., Pasadena, CA 91125, USA
}
\affil{
$^6$UCO/Lick Observatory, Department of Astronomy and Astrophysics, University of California, Santa Cruz, CA 95064, USA
}
\affil{
$^7$Department of Physics and Astronomy, Georgia State University,
  25 Park Place, Suite 605, Atlanta, GA 30303, USA
}
\affil{
$^8$Fermi National Accelerator Laboratory, P.O. Box 500, Batavia, IL 60510, USA 
}
\affil{
$^9$University of Science and Technology of China, Hefei 230026, China
}
\affil{
$^{10}$Nanjing Institute of Astronomical Optics \& Technology, National Astronomical Observatories, Chinese Academy of Sciences, Nanjing 210042, China
}

\begin{abstract}
We report the first hypervelocity star (HVS) discovered from the LAMOST 
spectroscopic survey. It is a B-type star with a heliocentric radial velocity 
about 620$\kms$, which projects to a Galactocentric radial velocity component 
of $\sim$477$\kms$. With a heliocentric distance of $\sim$13 kpc and an 
apparent magnitude of $\sim$13 mag, it is the nearest bright HVS 
currently known. With a mass of $\sim$9$M_\odot$, it is one of the three 
most massive HVSs discovered so far.
The star is clustered on the sky with many other known HVSs, 
with the position suggesting a possible connection to Galactic center 
structures. With the current poorly-determined proper motion, a Galactic 
center origin of this HVS remains consistent with the data at the 
$1\sigma$ level, while a disk run-away origin cannot be excluded. 
We discuss the potential of the LAMOST survey to 
discover a large statistical sample of HVSs of different types. 
\end{abstract}


\section{Introduction}

Hypervelocity stars (HVSs) are stars with velocities that exceed the escape
velocity of the Galaxy. They were first predicted by \citet{Hills88},
as a consequence of the tidal disruption of tight binary stars by the central
massive black hole (MBH) of the Galaxy. Since the first discovery of an HVS 
\citep{Brown05}, around 20 HVSs have been found 
\citep[][]{Edelmann05,Hirsch05,Brown06a,Brown06b,Brown07a,Brown07b,Brown12}. 
Here we report the discovery of the first HVS in the Large
Sky Area Multi-Object Fiber Spectroscopic Telescope (LAMOST) survey.

Besides the Hills mechanism, HVSs may also be produced by the interaction
between single stars and an intermediate-mass black hole inspiralling towards
the central MBH \citep[][]{Yu03}, as well as the tidal disruption of dwarf 
galaxies \citep{Abadi09}. Some HVSs may be runaway stars \citep{Blaauw61}, e.g.,
as the surviving companion stars in the white dwarf + helium star channel 
of Type Ia supernovae \citep{Wang09}, or from interactions among multiple stars
\citep[e.g.,][]{Gvaramadze09,Tutukov09}.

HVSs provide a unique probe for a wide range of Galactic science
\citep{Kenyon08}, on scales from a few pc (near the central MBH) to 
$\sim 10^5$ pc (the Galactic halo). The spatial and velocity distributions,
as well as the detection frequencies of HVSs, can be used to test the ejection 
mechanisms. If HVSs are ejected from the Galactic center (GC), 
the number density, velocity, and stellar type distributions of 
HVSs can reveal the environment around the central MBH and the stellar mass 
distribution near the MBH 
\citep[e.g.,][]{Brown06a,Kollmeier07,Lu07,Kollmeier10}.
The sky distribution of HVSs suggests a connection to the S stars in the two
disks near the central MBH \citep[e.g.,][]{Lu10}, which may provide
clues to the MBH growth \citep{Bromley12}. The trajectories of HVSs can 
also be used to probe the shape of the dark matter halo of the Galaxy 
\citep[][]{Gnedin05,Yu07}.

For all of the above applications, it is desirable to assemble a large, 
statistical sample of HVSs. The LAMOST survey has this potential, as described 
below.  In this paper, we report the first HVS discovered in the internal
Data Release 1 (DR1) of LAMOST.  In \S~2, we provide a brief 
description of the data, and then focus on the properties of the HVS and 
discuss the implications. Finally, in \S~3, we summarize the results and 
forecast the prospects of further HVSs discoveries from the LAMOST survey.

\section{The LAMOST Survey and Its First HVS}

\subsection{Data}

LAMOST is a 4m Schmidt telescope (now named the Guo Shoujing
Telescope) at the Xinglong Observing Station of the National Astronomical
Observatories of China. It is equipped with 4000 optical fibers in the
focal plane, taking spectra with resolution $R=\lambda/\Delta\lambda=1800$.
Within the LAMOST spectroscopic survey \citep{Cui12,Zhao12}, the LAMOST 
Experiment for Galactic Understanding and Exploration (LEGUE; \citealt{Deng12})
 aims to take $\sim$8 million stellar spectra for targets covering 
16,000 square degrees of sky over the course of five years. The 
target-selection criteria required to achieve various science goals can be 
found in \citet{Carlin12}, \citet{Chen12}, \citet{Yang12}, and \citet{Zhang12}.

The internal DR1 of LAMOST spectra and stellar parameters includes  
spectra obtained during a Pilot Survey (from October 2011 to June 2012; 
\citealt{Luo12}) and the first year of the regular survey (from September 
2012). Most of the observed stars are brighter than $r=16$. We performed a 
systematic search for HVSs in the catalog including stars of A-type or earlier,
and one of the $\sim 10^5$ stars turned out to be a HVS (hereafter, denoted 
LAMOST-HVS1).

\begin{deluxetable}{ll}
\tablecaption{\label{tab:tab1}
Properties of LAMOST-HVS1}
\tablehead{
& J091206.52+091621.8
}
\startdata
Position ($J$2000)        
& $(\alpha,\, \delta)=(138^\circ .027199,\, 9^\circ .272725)$ \\
& $(l,\, b)=(221^\circ .099564,\, 35^\circ .407261)$\\  
Magnitudes      & $g=12.91$ \, $r=13.22$ \, $i=13.50$ \\
                & $B=12.96$ $V=13.06$ $J=13.36$ \\
                & $H=13.43$ $K=13.53$\\
Distance        & $13.4\pm 2.2$ kpc (Heliocentric)\\
                & $19.4\pm 2.1$ kpc (Galactocentric)\\
Radial Velocity & $v_{r\odot}=620\pm 10 {\rm \,km\, s^{-1}}$ \\
                & $v_{rf}=477 \pm 10 {\rm \,km\, s^{-1}}$    \\
Proper Motion   & $(\mu_\alpha\cos\delta\, ,\,\,\,\,\,\,\,\,\, \mu_\delta)$ \\
\,\,\,(${\rm mas\, yr^{-1}}$)
                & $(-4.0\pm0.7,\, -4.9\pm 1.2)$ [UCAC4]\\
                & $(-2.5\pm1.9,\, -1.2\pm 1.9)$ [PPMXL]\\
                & $(\,\,\,\,\,0.9\pm1.9,\,\,\,\,\,0.9\pm 1.9)$ [cPPMXL]\\
Spectral Type & B   \\
$T_{\rm eff}$ & $ (2.07 \pm 0.12)\times 10^4$K       \\
$\log [g/({\rm cm\, s^{-2}})]     $ & $ 3.67 \pm 0.19$ \\
$[{\rm Fe/H}]$      & $-0.13\pm 0.07$                  \\
Mass & $9.1\pm 0.7 M_\odot$
\enddata

\end{deluxetable}

\subsection{Properties of LAMOST-HVS1}

The star (J091206.52+091621.8; LAMOST-HVS1) has been observed twice by LAMOST, 
separated
by about 70 days (December 23, 2012 and March 5, 2013). It is a bright star
with magnitude around 13. The radial velocities at the two epochs from 
spectral fitting are consistent with each other within the uncertainties, and 
therefore there is no evidence for it being a close binary system. 

The measured heliocentric radial velocity, $v_{r\odot}=620\pm 10\kms$, 
translates to a Galactocentric radial component $v_{rf}=477 \pm 10 \kms$, 
according to
\begin{equation}
v_{rf}=v_{r\odot} + U_0\cos l \cos b + (V_{LSR}+V_0)\sin l \cos b + W_0\sin b,
\end{equation}
where we adopt $V_{LSR}=250\kms$ for the velocity of the local standard of rest
 (LSR) \citep{Reid09,McMillan10} and $(U_0, V_0, W_0)=(11.1, 12.24, 7.25)\kms$ 
for the peculiar motion of the Sun with respect to the LSR \citep{Schonrich10}.
The star has proper motion measurements, 
$(\mu_\alpha\cos\delta,\, \mu_\delta)
=(-4.0\pm0.7,\, -4.9\pm1.2)\, {\rm mas\, yr^{-1}}$
in the UCAC4 catalog \citep{Zacharias13}
and 
$(-2.5\pm1.9,\, -1.2\pm 1.9)\, {\rm mas\, yr^{-1}}$ in the PPMXL catalog 
\citep{Roeser10}.
We find that quasars within two degrees around the star in the PPMXL 
catalog have a net proper motion, and if we correct for this systematic offset,
the proper motion of the star becomes $(0.9\pm1.9,\, 0.9\pm 1.9)\, 
{\rm mas\, yr^{-1}}$ (listed as cPPMXL in Table~\ref{tab:tab1}). 
A more accurate determination is clearly desirable.

\begin{figure}
\epsscale{1.1}
\plotone{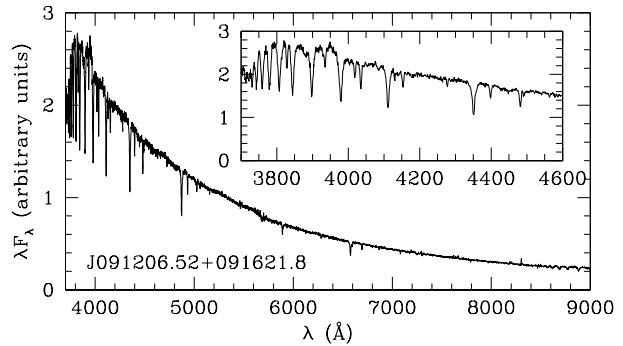}
\caption[]{
\label{fig:spec}
Spectrum of LAMOST-HVS1 taken with the Guo Shoujing Telescope. Shown in the 
inset is a zoomed-in view of the blue end of the spectrum.
}
\end{figure}

\begin{figure}
\epsscale{1.1}
\plotone{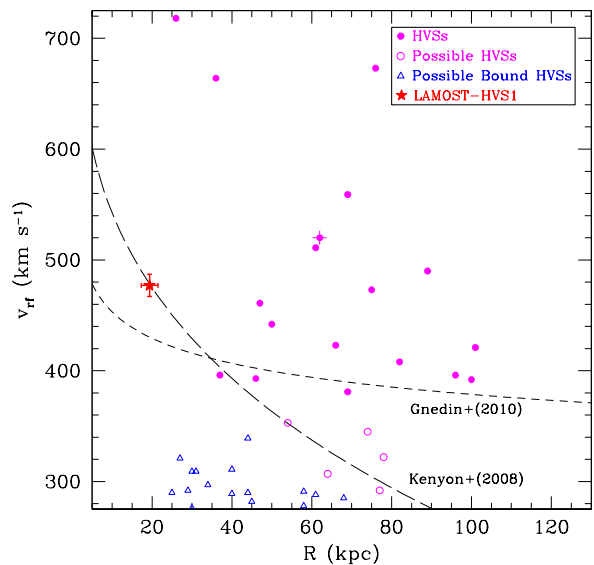}
\caption[]{
\label{fig:VR}
Galactocentric radial velocity of known/possible/bound HVSs \citep{Brown12}
and LAMOST-HVS1 versus the Galactocentric distance. The short and dashed 
curves are escape velocities from two models of Galactic potential 
\citep{Kenyon08,Gnedin10}, and the difference illustrates the current 
uncertainties in the models. The filled circle with a cross is HVS 
HE 0437-5439, LAMOST-HVS1's near twin (see text).
}
\end{figure}

Figure~\ref{fig:spec} shows the low-resolution LAMOST spectrum of the star. 
The basic stellar atmospheric parameters, listed in Table~\ref{tab:tab1}, are 
obtained by fitting the spectrum with the University of Lyon Spectroscopic 
analysis Software 
(ULySS\footnote{http://ulyss.univ-lyon1.fr/}; \citealt{Koleva09}). It
has effective temperature $T_{\rm eff}\simeq 2\times 10^4 {\rm K}$, surface
gravity $\log[g/({\rm cm\, s^{-2}})]\simeq 3.7$, and slightly sub-solar 
metallicity, $[{\rm Fe/H}]\simeq -0.13$.  The presence of strong helium lines 
indicates that it is an early B-type star.

To determine the distance and mass of LAMOST-HVS1, we make use of the list of 
masses and luminosities of B-type stars compiled by \citet{Hohle10}, based on 
2MASS photometry and Hipparcos parallax. By matching the values of temperature 
and surface gravity to the range derived from spectral fitting of the 
LAMOST-HVS1, we infer its spectral type to be between B1 and B2.5. While the 
luminosity class is not well-constrained, it is correlated with the spectral
type. For example, the star could be B1I, B2IV, or B2.5V. The degeneracy
leads to only a small luminosity variation among such stars, which translates 
to a relatively well-constrained distance, $13.4\pm 2.2$ kpc. With 8 kpc 
adopted for the Sun's distance to the GC, the Galactocentric 
distance of LAMOST-HVS1 is calculated to be $R=19.4\pm 2.1$ kpc. The mass of 
LAMOST-HVS1 is inferred to be $9.1\pm 0.7 M_\odot$. 

Interestingly, LAMOST-HVS1 appears to be almost a twin to HE 0437-5439 
(a.k.a HVS 3; e.g., \citealt{Edelmann05,Bonanos08,Przybilla08}), which is also 
a $\sim$9$M_\odot$ B-type star, with similar temperature and surface gravity. 
The Galactic position of LAMOST-HVS1 shares some similarities with HD 271791 
\citep{Heber08}, a $11\pm1 M_\odot$ B-giant stars established to be a run-away
star with velocity similar to those of hypervelocity stars. HD 271791 is 
$21.8\pm 3.7$ kpc away from the GC and $-10.4\pm 2.0$ kpc below the disk 
plane \citep{Heber08}, while for LAMOST-HVS1 those are $19.4\pm 2.1$ kpc and 
$7.8\pm 1.3$ kpc. Together, the above three stars make the most massive HVSs 
discovered so far, and LAMOST-HVS1 is the nearest one.

With the velocity and distance determined, Figure~\ref{fig:VR} places 
LAMOST-HVS1 in the $v_{rf}$--$R$ plane, and compares it to the known HVSs, 
as well as possible HVSs and possible bound HVSs, as listed in \citet{Brown12}.
 Clearly, LAMOST-HVS1 is the nearest HVS discovered so far. Following 
\citet{Brown12}, we also plot two curves (long and short dashed) of escape 
velocities, based on the Galactic potential models of \citet{Kenyon08} and 
\citet{Gnedin10}, respectively. At $R\sim 20$ kpc, the velocity 
$v_{rf}=477\kms $ of LAMOST-HVS1 is above the escape velocity in the 
\citet{Gnedin10} model and falls almost on top of the model curve from 
\citet{Kenyon08}. Note that including the proper motion contribution (using
the PPMXL one for the most conservative estimate) implies a total velocity
of $545\pm 54\, {\rm km\, s^{-1}}$, with the uncertainties from distance,
proper motion, and radial velocity. This establishes LAMOST-HVS1's identity 
as an HVS.

\subsection{Galactic Center Origin?} 

\begin{figure}[t]
\epsscale{1.2}
\plotone{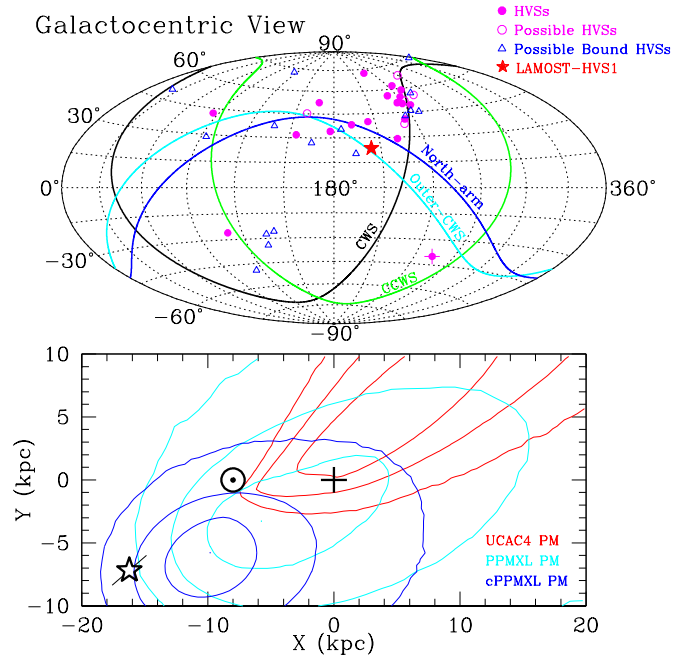}
\caption[]{
\label{fig:sky_disk}
{\it Top:} 
Galactocentric view of the sky distribution of HVSs in the Galactic coordinate 
system, and the position of LAMOST-HVS1. The positions of stars are shifted to 
match what would be seen by an observer at the Galactic center. The great 
circles correspond to structures of stellar distribution near the massive 
black hole at the Galactic center (see text). The filled circle with a cross
is HE 0437-5439, LAMOST-HVS1's twin (see text).
{\it Bottom:}
Distribution of ejection position, if LAMOST-HVS1 originated from the Galactic 
plane. The contours indicate the 1$\sigma$, 2$\sigma$, and 3$\sigma$ ranges 
(color-coded for using UCAC4, PPMXL, and cPPMXL proper motions, respectively; see the text), and 
the loose constraint is mainly a consequence of the large uncertainty in the 
proper motion. The projected position of LAMOST-HVS1 is shown as the star, 
with the line denoting its 1$\sigma$ uncertainty. The Sun's position and the 
Galactic center are marked with a $\odot$ and a cross, respectively.
}
\end{figure}

We now explore to what extent the data constrain the origin of LAMOST-HVS1.

\citet{Lu10} and \citet{Zhang13} demonstrate that, under the tidal-disruption 
scenario, the spatial distribution of HVSs can track that of the progenitors. 
\citet{Lu10} indeed find that most of the discovered HVSs, if viewed from the 
GC, show spatial distributions near the great circles connecting to the planes 
of the clockwise-rotating young stellar (CWS) disk and the northern arm of the 
mini-spiral or the outer wrapped part of the CWS disk, supporting the GC 
origin of HVSs (but see also \citealt{Pawlowski13}).

Following \citet{Lu10}, the top panel of Figure~\ref{fig:sky_disk} shows the 
sky distribution of the 
known and possible HVSs and the position of LAMOST-HVS1, {\it as viewed from 
the GC}. The great circles correspond to different stellar structures in the
GC: the CWS, the outer wrap of the CWS (Outer-CWS), the Northern arm 
(North-arm), and the counter-clockwise stellar disk (CCWS). Interestingly, 
LAMOST-HVS1 falls into the clustered region defined by most other known HVSs. 
It is closest to the Outer-CWS great circle and also close to the one for the 
North-arm. This seems to support its GC origin, and suggests that its 
progenitors were associated with the CWS or North-arm. 

In the top panel of Figure~\ref{fig:sky_disk}, LAMOST-HVS1's twin, 
HE 0437-5439, lies near the CCWS
circle. But it is also possible to connect to the Outer-CWS, similar to 
LAMOST-HVS1. HE 0437-5439 has been proposed to have been produced from the 
Large Magellanic Cloud \citep[][]{Edelmann05,Bonanos08,Przybilla08}, but
with well-measured proper motion \citet{Brown10} conclude that it is more 
likely a compact binary ejected from the GC, which later evolved into a blue 
straggler. 

The proper motion of LAMOST-HVS1 is not well-measured.  We perform a Monte 
Carlo simulation, by accounting for the uncertainties in the proper motion 
(for the PPMXL, cPPMXL, and the UCAC4 value), distance, and radial velocity 
(all assumed to be Gaussian), in order to consider the implications of the 
velocity vector for the origin of LAMOST-HVS1. Obviously, the uncertainties 
are dominated by those in the proper motion.

From the star's current position, we go along the opposite direction of the 
velocity vector and perform a full orbit integration with the Galactic 
potential model in \citet{Kenyon08} to derive the intercept position in the 
disk plane. This would be the point of origin for the HVS if it were ejected 
from the plane. The Monte Carlo simulation shows 
that the most likely intercept depends on which proper motion measurement 
is used (see the bottom panel of Figure~\ref{fig:sky_disk}). The GC is within 
the $\sim 1\sigma$ ($<1\sigma$) range of the distribution of the intercept 
positions with the UCAC4 (PPMXL) proper motion adopted. If we use the PPMXL
proper motion corrected for the offset of quasars, the data then favor a disk
origin, with the most likely intercept lies close to the Perseus spiral arm.
An accurate measurement of the proper 
motion will be a key to determining the origin of LAMOST-HVS1. If the GC
origin holds, the flight time ($\sim$32 Myr) for LAMOST-HVS1 to reach its 
current position would be comparable to its lifetime (estimated
to be around 30--40 Myr). If the ejection is delayed by a few million years
or more since the formation \citep[e.g.,][]{Brown12}, there would be some 
tension between the delay + flight time and the lifetime. Therefore it may 
be either a massive star ejected 
directly from the GC (with properties in broad agreements with predictions by 
\citealt{Zhang13}), or a blue straggler from similar processes as HE 0437-5439.

We cannot rule out a disk (run-away) origin of LAMOST-HVS1, especially if we 
use the cPPMXL proper motion. As mentioned before, the position of LAMOST-HVS1 
reminds us of the B-type giant HD 271791, which is about 10 kpc from the disk 
plane. Its proper motion measurement clearly favors a disk origin, establishing
 it as a hyper-velocity run-away star \citep{Heber08}. \citet{Bromley09} show 
that high speed stars near the disk should mostly be disk run-away stars, and 
HD 271791 is therefore an example of unbound run-away stars. For LAMOST-HVS1,
a disk run-away origin would greatly alleviate the potential tension between 
flight time 
and lifetime of the star as in the GC-origin scenario. A better proper motion 
determination for LAMOST-HVS1 will help to show whether the possibility of 
a disk run-away star holds. 

Given that LAMOST-HVS1 is the brightest example of known HVSs and its clear 
similarity to HE 0437-5439 and HD 271791, it will be extremely interesting to 
conduct 
a more detailed study. If indeed it comes from the GC, we expect a proper 
motion around 3--4.5 ${\rm mas\, yr^{-1}}$, 
well in reach of current and near-future data. High-resolution spectroscopic 
study will reveal its chemical abundance pattern and rotation velocity. We 
plan to perform such follow-up observations and investigations.

\section{Summary and Discussion}

\begin{figure}
\epsscale{1.1}
\plotone{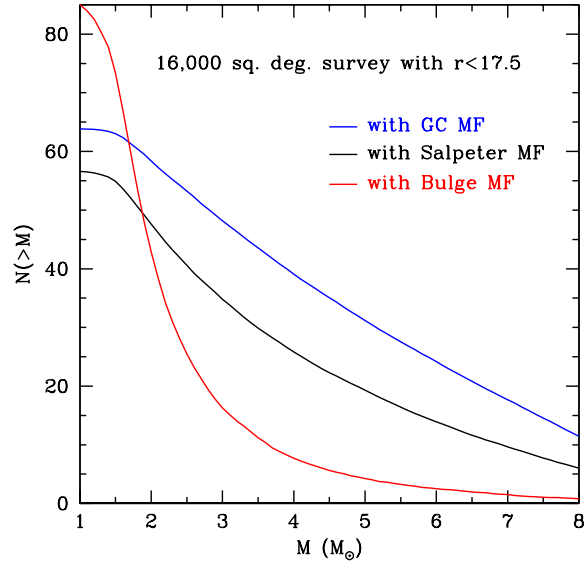}
\caption[]{
\label{fig:forecast}
Prospect of discovering (unbound) HVSs from the LAMOST survey, based
on the plan of a 16,000 square degree survey with a limiting magnitude 
$r=17.5$.  The curves represent the numbers of HVSs that LAMOST may discover 
for three cases of the stellar MF. 
}
\end{figure}

We present the first HVS discovered from the LAMOST survey, a B-type star with 
a mass of $\sim$9$M_\odot$, located at a Galactocentric distance of $\sim$19 
kpc with a Galactocentric radial velocity component of $\sim$477$\kms$, based 
on the measured radial velocity. It is the nearest, and one of 
the three most massive HVSs discovered so far. 

LAMOST-HVS1 is clustered with most other known HVSs on the sky. Its proximity 
to the great circles corresponding to stellar structures around the central 
MBH suggests a GC origin. A more accurate proper motion measurement, 
achievable in the near future, will pin down its origin.
 
LAMOST-HVS1 signals the start of the HVS discovery effort from the LAMOST 
survey. The survey has the potential to discover a large number of HVSs. 
Figure~\ref{fig:forecast} shows a conservative forecast with a Monte Carlo
method for the plan of a 
16,000 square degree survey with a limiting magnitude of $r=17.5$ (the final 
survey may be deeper), following a LEGUE halo target selection with higher 
priority on bluer stars \citep{Carlin12}. The forecast assumes a GC origin of 
HVSs and is normalized by using the space density 
$0.077(R/{\rm kpc})^{-2} \,\,\,
{\rm kpc}^{-3}$ of $3M_\odot<M<4M_\odot$ HVSs in \citet{Brown07b}. A maximum
mass of $10 M_\odot$ is adopted in computing the cumulative count.
Three cases are considered: a Salpeter mass function (MF; 
\citealt{Salpeter55}), a Galactic bulge MF \citep{Mezger99}, and a GC 
MF \citep{Lu13}. The forecast implicitly assumes that in the mass range
considered, stars have identical binary fractions, distribution of binary 
orbital separations, and MBH ejection velocities, which may not be true in
detail but serves our purpose of estimate.

Since the LAMOST survey does not preselect B-type stars as targets for 
spectroscopic observations to reduce the contamination by the large number of 
halo stars, it can discover HVSs of various stellar types, from B to G. The 
total number of HVSs down to $M=1M_\odot$ with the above survey parameters is 
in the range of $\sim$56--85. In addition, we also expect to find a large 
number of bound HVSs and possibly binary HVSs. Such a large statistical sample 
of HVSs of different types would enable a wide range of investigations
to elucidate the nature of HVSs, and to constrain Galactic structure.

\acknowledgments
Z.Z. thanks Youjun Lu, Qingjuan Yu, and Ben Bromley for helpful discussions.
We thank the referee for a prompt and constructive report.
This work is supported by the U.S. National Science Foundation (NSF), through 
grant AST-0937523. T.C.B. acknowledges partial support from U.S. NSF grant 
PHY 08-22648. P.G. thanks the U.S. NSF for grant AST-1010039. Guoshoujing 
Telescope (the Large Sky Area Multi-Object Fiber Spectroscopic Telescope 
LAMOST) is a National Major Scientific Project built by the Chinese Academy 
of Sciences. Funding for the project has been provided by the National 
Development and Reform Commission. LAMOST is operated and managed by the 
National Astronomical Observatories, Chinese Academy of Sciences.

\end{document}